\documentclass[aps,prd,twocolumn,amsmath,showpacs,amssymb,superscriptaddress,nofootinbib]{revtex4-1}
\usepackage{graphics,graphicx}
\usepackage{amsmath,amssymb}
\usepackage{bbold}
\usepackage{color}
\usepackage{hyperref}
\usepackage[normalem]{ulem}  

\begin{document}
\title{ Thermodynamics of the strange baryon system from a coupled-channels analysis and missing states }
\date{\today}
\author{C\'esar Fern\'andez-Ram\'irez}
\affiliation{Instituto de Ciencias Nucleares, Universidad Nacional Aut\'onoma de  M\'exico, 04510 Mexico City, Mexico}
\author{Pok Man Lo}
\affiliation{Institute of Theoretical Physics, University of Wroclaw,
PL-50204 Wroc\l aw, Poland}
\author{Peter Petreczky}
\affiliation{Physics Department, Brookhaven National Laboratory, Upton, NY 11973, USA}

\begin{abstract}

We study the thermodynamics of the strange baryon system by using an S-matrix formulation of statistical mechanics. 
For this purpose, we employ an existing coupled-channel study involving $\bar{K} N$, $\pi \Lambda$, and $\pi \Sigma$ interactions in the $S=-1$ sector.
A novel method is proposed to extract an effective phase shift due to the interaction, 
which can subsequently be used to compute various thermal observables via a relativistic virial expansion.
Using the phase shift extracted,
we compute the correlation of the net baryon number with strangeness ($\chi_{BS}$) for an interacting hadron gas.
We show that the S-matrix approach, which entails a consistent treatment of resonances and 
naturally incorporates the additional hyperon states which are not listed by the Particle Data Group, 
leads to an improved description of the lattice data over the hadron resonance gas model.
\end{abstract}

\maketitle

\section{ Introduction }

It was conjectured a long time ago that thermodynamics of hadrons 
can be understood in terms of the hadron resonance gas (HRG) model~\cite{Hagedorn:1965st}. 
The essence of this model is that interacting hadron gas can be replaced by an uncorrelated gas of hadrons and hadronic resonances and, as a first approximation, resonances are treated as zero-width states. 
Lattice QCD (LQCD) calculations confirm that indeed the thermodynamics below
the chiral transition temperature $T<T_c \simeq 155$ MeV \cite{Borsanyi:2010bp,Bazavov:2011nk,Bhattacharya:2014ara} 
can be understood in terms of the HRG model.
The model appears to describe the pressure and the trace anomaly calculated on the lattice
\cite{Karsch:2003vd,Huovinen:2009yb,Borsanyi:2010bp,Bazavov:2014pvz,Bazavov:2017dsy}.
Fluctuations, $\chi_n^X$ and correlation $\chi_{nm}^{XY}$ 
of conserved charges defined as the derivatives of the pressure with
respect to chemical potentials $\mu_{X,Y}$,
\begin{eqnarray}
&
\displaystyle
\chi_n^X=T^n \frac{\partial^n (p(T,\mu_X)/T^4)}{\partial \mu_X^n}|_{\mu_X=0},\\[2mm]
&
\displaystyle
\chi_{nm}^{XY}=T^{n+m} \frac{\partial^{n+m} (p(T,\mu_X,\mu_Y)/T^4)}{\partial \mu_X^n \partial \mu_Y^m}|_{
\mu_X=0,\mu_Y=0},
\end{eqnarray}
are also reasonably well described by the HRG model~\cite{Karsch:2003zq,Huovinen:2009yb,Borsanyi:2011sw,Bazavov:2012jq}.
Here the conserved charges $X,Y=B,~Q,~S$, correspond
to baryon number, electric charge and strangeness, respectively.
In recent years the HRG model was extended to include
the effects of excluded 
volume (see e.g., Refs. \cite{Begun:2012rf,Albright:2015uua,Vovchenko:2016rkn,Chatterjee:2017yhp,Samanta:2017yhh}) 
as well as finite resonance width (see e.g., Refs. \cite{Blaschke:2003ut,Jakovac:2013iua}).

Around the time of the original Hagedorn's proposal
a more systematic approach to study thermodynamics of hadrons was proposed
by Dashen, Ma, and Bernstein-- the S-matrix formulation of statistical mechanics \cite{Dashen:1969ep}. 
It is an extension of the usual virial expansion to the relativistic case. When used in conjunction with the empirical 
phase shifts from scattering experiments, 
this approach offers a model-independent way to consistently 
incorporate the effects of hadronic interactions, including the appearance of broad resonances and purely repulsive channels.
The analysis of Venugopalan and Prakash \cite{Venugopalan:1992hy} along these lines showed that, for the pressure, there is
a large cancellation of different nonresonant repulsive and attractive contributions and therefore
the pressure to a fairly good approximation can be understood solely in terms of resonances alone, 
thus justifying the use of the HRG model \cite{Venugopalan:1992hy}. 
For a more recent analysis, see Ref.~\cite{Wiranata:2013oaa}.
To what extent this simplifications can be justified
for the fluctuations and correlations of conserved charges remains to be seen. 
Recently, the S-matrix approach has been applied to analyze the LQCD 
result on the baryon electric charge correlation $\chi_{11}^{BQ}$~\cite{Lo:2017lym}.
The former observable is particularly sensitive to the interaction between pions and nucleons.
It was found that the use of the effective density of states constructed from the phase shifts 
of a partial wave analysis (PWA) leads to an improved description of the LQCD result 
up to a temperature $T\approx 160$ MeV over that of the HRG model.
The source of the improvement in the quantitative description of the LQCD result within the S-matrix approach is twofold. 
First, the inclusion of nonresonant, often purely repulsive, 
channels yields an important contribution at low invariant masses. 
Second, a consistent treatment of the interactions is pivotal in channels with broad resonances. 
For such a resonance, the thermal contribution can be significantly reduced relative to the HRG prediction owing to 
the fact that a substantial part of the effective density of states is found at large masses, 
which are suppressed by the Boltzmann factors. 

The S-matrix approach was also considered to describe 
the pressure of the nucleon gas \cite{Huovinen:2017ogf}. Here the nonresonant and dominantly repulsive
nucleon-nucleon interactions are important.
Based on these considerations an HRG model
with repulsive mean field was constructed and the higher-order baryon number fluctuations were 
calculated. It was found that repulsive mean field can describe the lattice results for the
higher-order baryon-number fluctuations \cite{Huovinen:2017ogf}. Here we note
that the effect of repulsive baryon-baryon
interactions was also modeled by excluded volume \cite{Vovchenko:2016rkn,Vovchenko:2017zpj}. This approach,
too, is able to describe higher-order baryon-number fluctuations. So there appears to be a consensus
that repulsive baryon-baryon interactions are important. We will return to this issue at
the end of the paper.

It is natural to expand the previous S-matrix studies involving nucleons to include also the strange baryons.
This problem demands a coupled-channel treatment 
as inelastic $(\bar{K} N, \pi \Lambda, \pi \Sigma)$ interaction sets in at rather low momentum.
Moreover, the available experimental data are insufficient to constrain individual scattering process and 
model extrapolations for poorly (if at all) measured amplitudes are hard to avoid.
A channel-by-channel analysis to describing the thermal system, as is done previously, would be rather inefficient.
In the hadron physics community, a multichannel S-matrix is usually constructed through a model of the amplitude 
that attempts to incorporate unitarity, analyticity, crossing symmetry, and any underlying symmetry of the given reactions
to describe the available data. The model S-matrix thus obtained can be used to 
study the thermal properties of the hadronic medium. 
In this paper we study the pressure of strange baryons in the S-matrix approach.
For this purpose we  introduce a robust method to extract an effective phase shift from a general coupled-channel PWA.
This phase shift encodes essential information about the scattering system and 
can be used to compute various thermal observables via a relativistic virial expansion.
As an application of the calculation scheme, we compute the pressure of baryons with strangeness one and
the second-order correlation of the net baryon number with strangeness ($\chi^{BS}_{11}$) for an interacting hadron gas.

We show that the proper treatment of resonances in the S-matrix approach 
leads to an improved description of the lattice data over the HRG model.
An important feature of our analysis is that the PWA has more resonances than
the list of three and four star resonances in the Particle Data Group (PDG) \cite{PDG}.
Thus, the S-matrix approach with the state-of-the-art PWA confirms an earlier conjecture that the description
of the lattice data on $\chi_{11}^{BS}$ requires additional baryon resonances 
not listed in the PDG \cite{Bazavov:2014xya,Alba:2017mqu}.

\section{ Effective phase shift for coupled-channel system }
\label{sec2}

\begin{figure*}[!ht]
\includegraphics[width=3.355in]{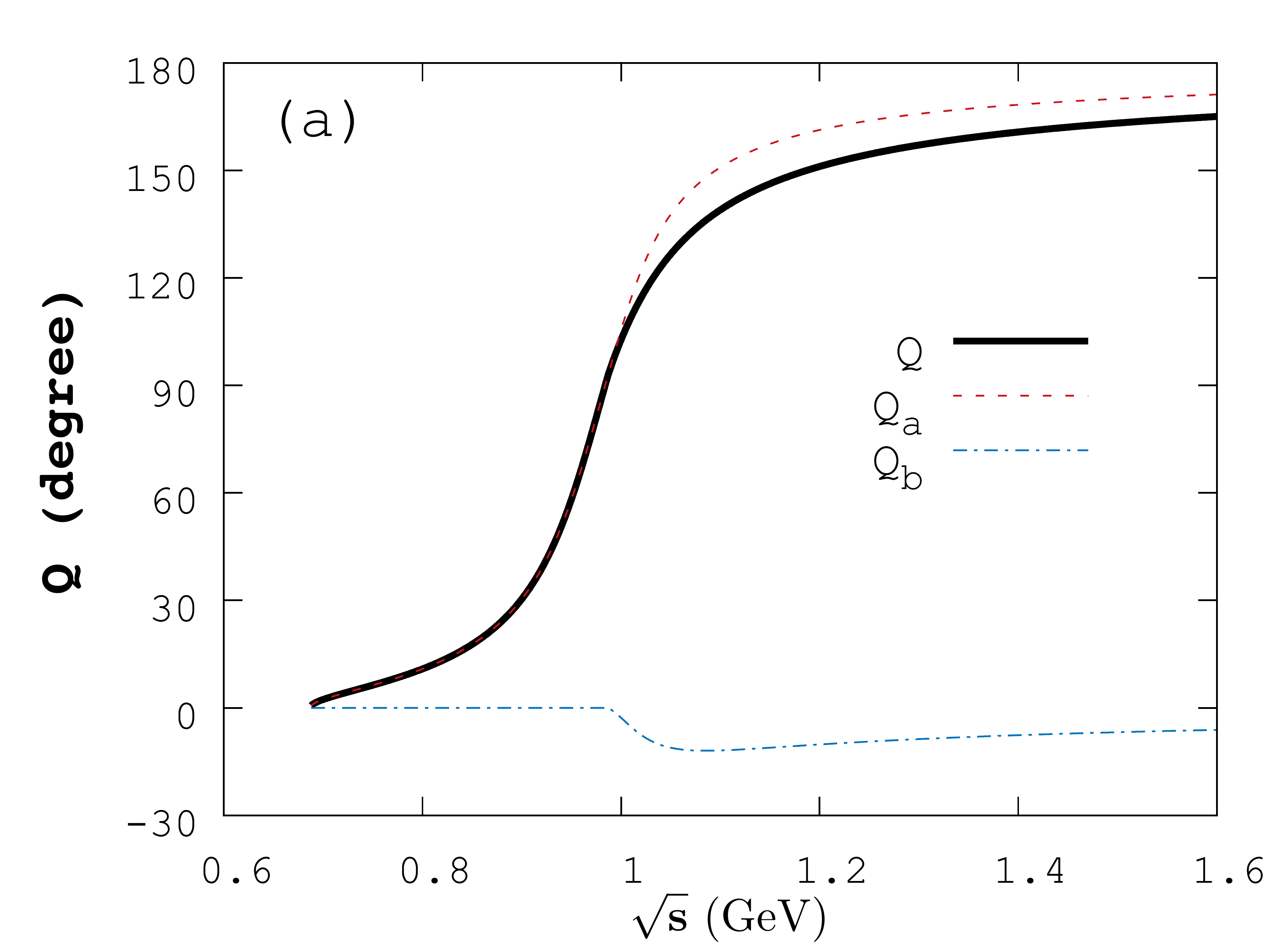}
\includegraphics[width=3.355in]{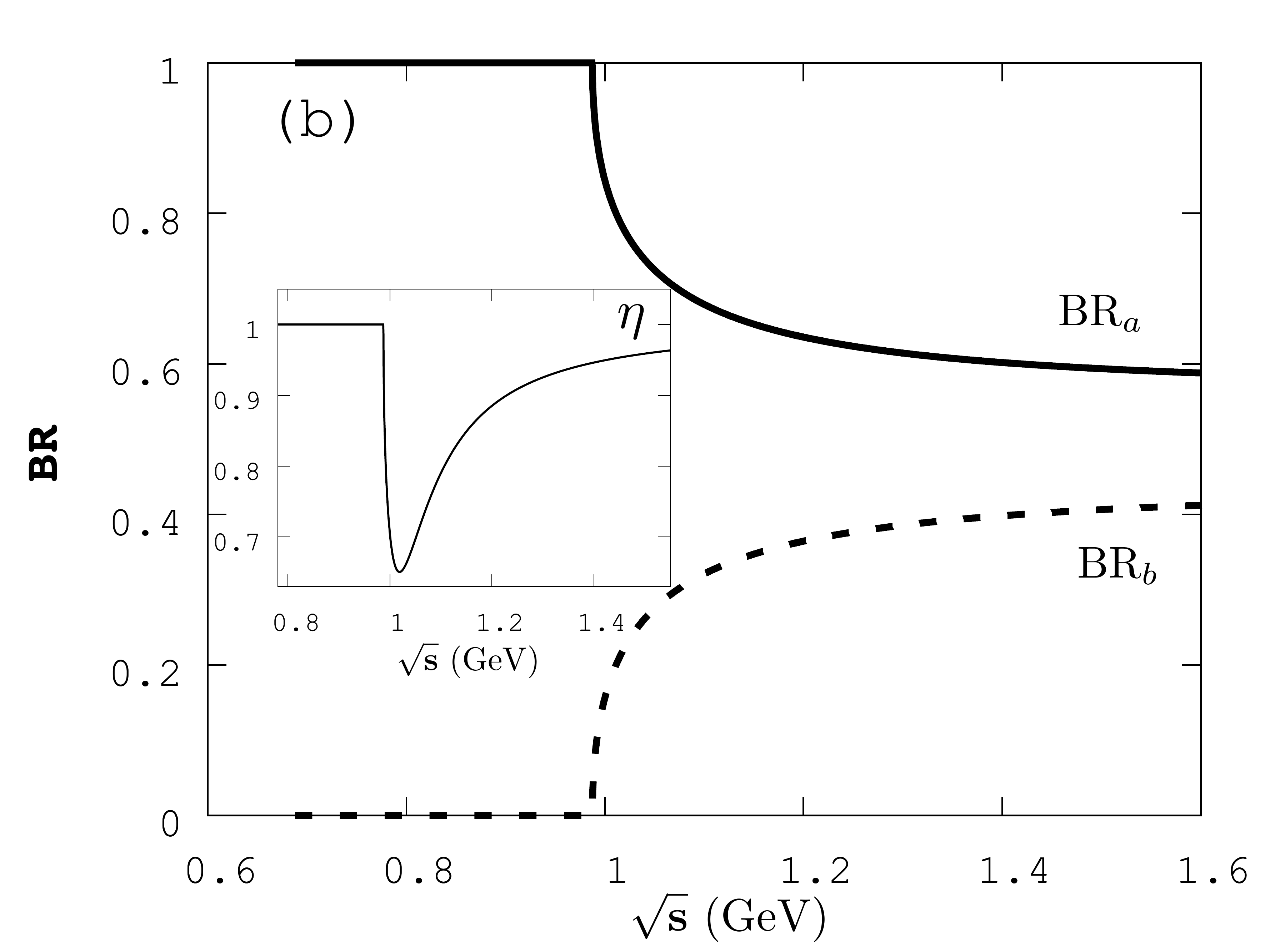}
  \caption{FIG.1.(a) Phase shifts {(Eq.~\eqref{eq:Qi})} and (b) branching ratios {(Eq.~\eqref{eq:branching})} obtained from a model describing the decay of a single resonance: $a_0(980)$
into two channels, $\pi \eta$ (channel $a$) and $K \bar{K}$ (channel $b$). 
The inset in the right panel shows the energy dependence of the inelasticity parameter $\eta$ {(Eq.~\eqref{eq:Qi})}.
  }
\label{fig:fig1}
\end{figure*}

\begin{figure*}[!ht]
\includegraphics[width=3.355in]{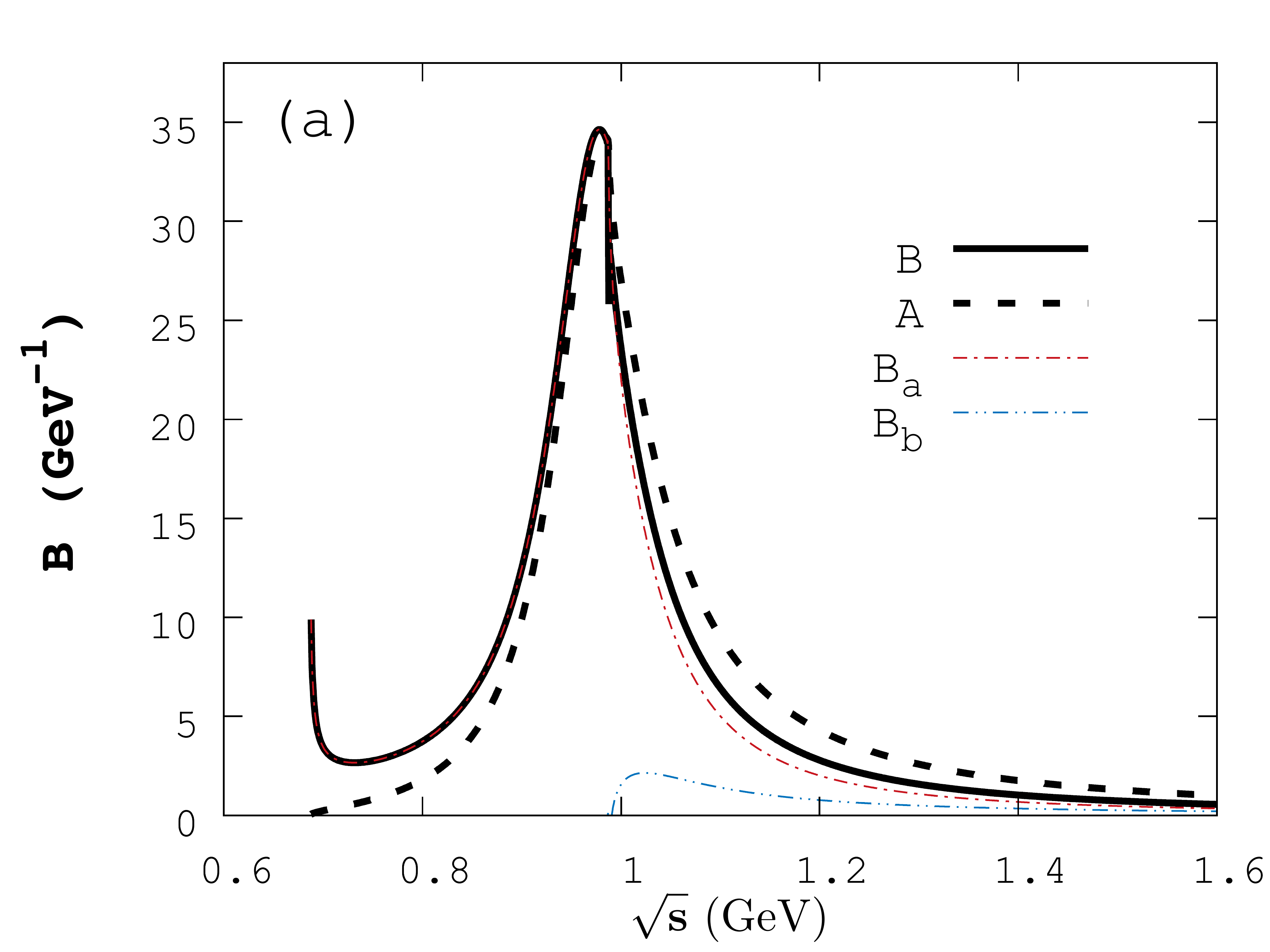}
\includegraphics[width=3.355in]{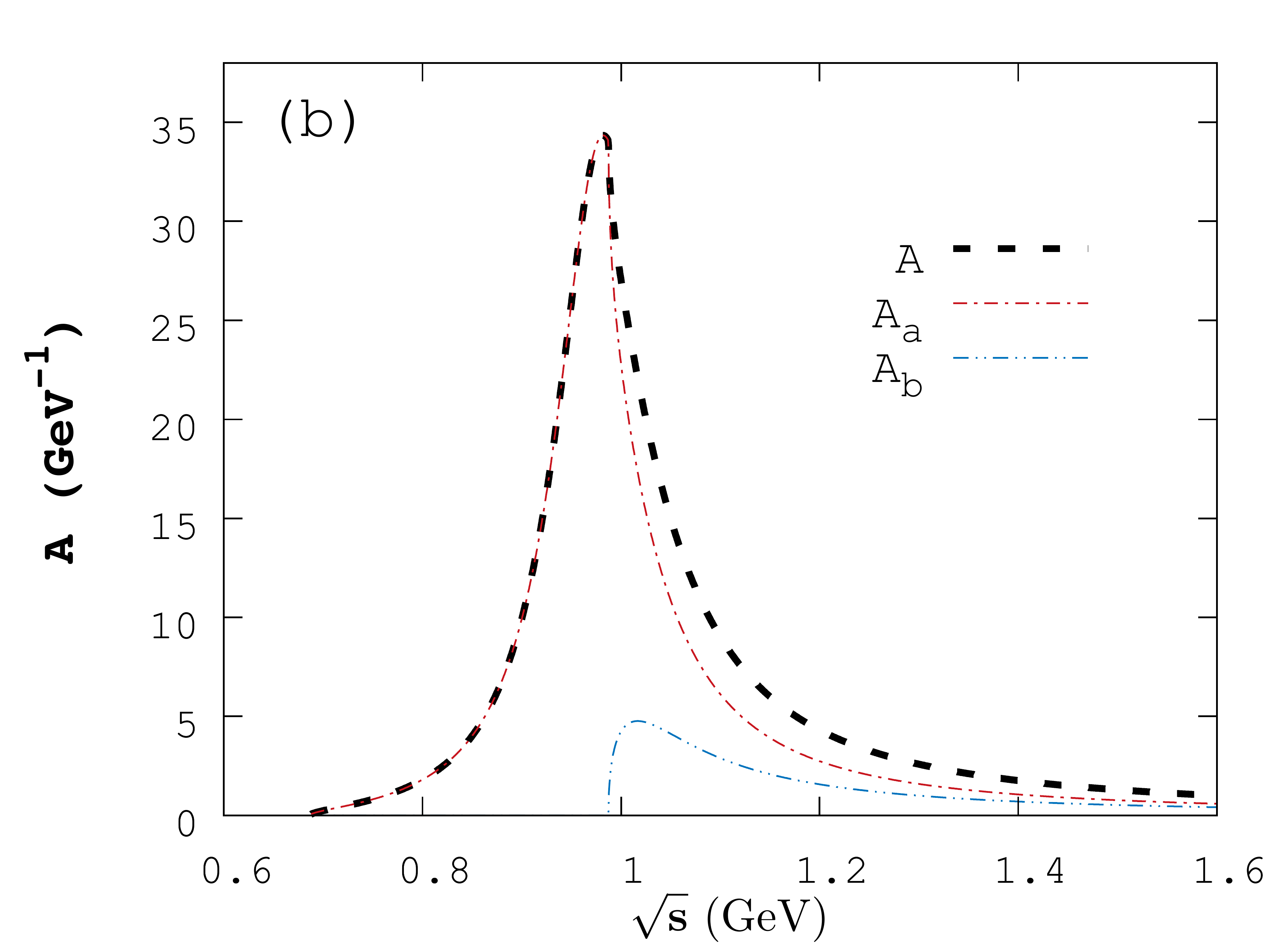}
\caption{The total and channel-specific effective (a) and standard (b) spectral functions defined in Eq.~\eqref{eq:bna}.}
\label{fig:fig2}
\end{figure*}

Before getting into the details of extracting an effective phase shift from a full-fledged coupled-channel PWA for the 
$S=-1$ baryons, we begin by introducing the method in a simpler setting: an elementary two-channel resonance decay model.

Consider the following fact of a unitary S-matrix $S$ (in a given partial wave), 

\begin{align}
  \begin{split}
    &S S^\dagger = \mathbb{1} \\
    \Rightarrow \, &{\rm det} S  \times {\rm det} S^\dagger = 1 \\
    \Rightarrow \, &\ln {\rm det} S  +  \ln {\rm det} S^\dagger = 0.
  \end{split}
\end{align}

\noindent Using $\ln {\rm det} S^\dagger = (\ln {\rm det} S )^*$, we see that unitarity dictates the quantity $(\ln {\rm det} S)$ to be purely imaginary.
This motivates the definition of a generalized phase shift function~\cite{Lo:2017sde,Suzuki:2008rp} $\mathcal{Q}$

\begin{align}
  \label{eq:psq}
  \begin{split}
    \mathcal{Q} &\equiv \frac{1}{2} \, \rm{Im} \left( \, {\rm tr} \,  \ln \, S \, \right) \\
    &= \frac{1}{2} \, \rm{Im}  \left( \, \ln \, {\rm det} \,S \, \right).
    \end{split}
\end{align}

\noindent The determinant operation makes this quantity invariant under any unitary rotation ($U$) of the S-matrix 

\begin{align}
 S \rightarrow U^\dagger S U.
\end{align}

In the single channel case $\mathcal{Q}$ reduces to the standard expression of a scattering phase shift. 
The physical meaning of this quantity in the general N-channel case can be clarified by studying a simple example.
Consider a single relativistic Breit-Wigner (B-W) resonance of mass $m_{\rm res}$ decaying via two channels.
The S-matrix can be parametrized as~\cite{Chung:1995dx,Haberzettl:2007ww,Kelkar:2008na}

\begin{align}
  \label{eq:example1}
  S({s}) &= \mathbb{1} + i \, \hat{T}({s}),
\end{align}

\noindent where, e.g., for the $l=0$ partial wave, 

\begin{align}
  \label{eq:example2}
  \begin{split}
    \hat{T}({s}) &= \frac{ -2 \sqrt{s} \, \gamma_{\rm res} }{ s - m_R^2 + i \, \sqrt{s} \, \gamma_{\rm res} }  \times \hat{t} \\
    \hat{t} &= \frac{1}{g_a^2 \, \phi_a + g_b^2 \, \phi_b} \left( \begin{array}{cc}
      g_a^2 \, \phi_a  &   g_a g_b \, \sqrt{\phi_a \phi_b} \\
 g_a g_b \, \sqrt{\phi_a \phi_b}  & g_b^2 \, \phi_b \end{array} \right).
  \end{split}
\end{align}

\noindent In this parametrization, $\sqrt{s}$ is the invariant mass, $g_{a}$ and $g_{b}$ are coupling constants (with $g_a^2+g_b^2=1$), and $\phi_{a}({s})$ and $\phi_{b}({s})$ are the relevant Lorentz-invariant phase space~\cite{Lo:2017sde}. 
For the two-body case it takes the generic form

\begin{align}
  \begin{split}
    \phi_2(s) &= \int \frac{d^3 p_1 d^3 p_2}{(2 \pi)^6} \frac{1}{4 E_1 E_2} \, \\
    & \times (2 \pi)^4 \delta(\sqrt{s}-E_1-E_2) \, \delta^{(3)}(\vec{p}_1+\vec{p}_2) \\
    &=\frac{q}{4 \pi \sqrt{s}}, 
  \end{split}
\end{align}

\noindent where

\begin{align}
    q &= \frac{1}{2} \sqrt{s} \, \sqrt{1-\frac{(m_1+m_2)^2}{s}} \sqrt{1-\frac{(m_1-m_2)^2}{s}},
\end{align}

\noindent and $m_1, m_2$ are the masses of the particles making up the channel. The (energy-dependent) total width of the resonance can be computed by

\begin{align}
  \gamma_{\rm res}({s}) = \gamma_0 \times \left( \, g_a^2 \, \phi_a + g_b^2 \, \phi_b  \, \right),\label{eq:gammaedependence}
\end{align}

\noindent with a width parameter $\gamma_0$.

A direct calculation shows that the phase shift function $\mathcal{Q}$ defined in Eq.~\eqref{eq:psq} is given by

\begin{align}
  \mathcal{Q}({s}) = \delta_{\rm res}({s}) = \tan^{-1} \frac{-\sqrt{s} \, \gamma_{\rm res}}{s-m_{\rm res}^2}.
\end{align}

\noindent We see that $\mathcal{Q}$ correctly recovers the phase shift of a relativistic B-W resonance~\cite{Friman:2015zua}.
This result is independent of the basis being used. In fact one can rewrite the model S-matrix in Eqs.~\eqref{eq:example1} and \eqref{eq:example2} as

\begin{align}
  S = U^\dagger \, S_d \, U
\end{align}

\noindent with  

\begin{align}
  \begin{split}
  S_d &= \left( \begin{array}{cc}
    e^{2 i \delta_{\rm res}({s})}  &  0 \\
    0  &  1 \end{array} \right), \\
  U &= \left( \begin{array}{cc}
      \cos\theta  &  \sin\theta \\
    -\sin\theta  &  \cos\theta \end{array} \right).
  \end{split}
\end{align}

\noindent This means that the ``observed'' S-matrix is related to a diagonal eigenmatrix $S_d$ (made up of eigenphases~\cite{Wigner:1955zz, Manley:1992yb}) by a rotation matrix whose elements can be related to the energy-dependent branching fractions:

\begin{align}
  \label{eq:branching}
  \begin{split}
    {\rm BR}_a \equiv \cos^2 \theta &= \frac{g_a^2 \, \phi_a}{ g_a^2 \, \phi_a + g_b^2 \, \phi_b }, \\
    {\rm BR}_b \equiv \sin^2 \theta &= \frac{g_b^2\, \phi_b }{ g_a^2 \, \phi_a + g_b^2 \, \phi_b }.
  \end{split}
\end{align}

\noindent Note that $\mathcal{Q}$ is invariant under a change of basis and is hence independent of these branching fractions.

To obtain other channel-specific quantities, we compare the model S-matrix with the general two-channel parametrization

\begin{align}
  S = \left( \begin{array}{cc}
    \eta \, e^{2  i  \mathcal{Q}_a}  &   i \sqrt{1-\eta^2} \, e^{i (\mathcal{Q}_a + \mathcal{Q}_{b})} \\
    i \sqrt{1-\eta^2} \, e^{i  (\mathcal{Q}_{a} + \mathcal{Q}_{b})}   &  \eta \, e^{2  i  \mathcal{Q}_{b}}
  \end{array} \right).
\end{align}

\noindent The inelasticity parameter $\eta$ and the channel phase shifts $\mathcal{Q}_{i=a,b}$'s can be extracted from , e.g., the diagonal elements of the S-matrix, via

\begin{align}
\label{eq:Qi}
  \begin{split}
    \mathcal{Q}_i &= \frac{1}{2} \, {\rm Im} \ln \, S_{ii}, \\
    \eta &=  e^{\, {\rm Re} \ln S_{ii}}.
  \end{split}
\end{align}

\noindent It also follows from Eq.~\eqref{eq:psq} that $\mathcal{Q} = \mathcal{Q}_a + \mathcal{Q}_b$. 

In Fig.~\ref{fig:fig1} we illustrate some of these quantities within the single resonance model.
The model describes the decay of the $a_0(980)$ resonance into $\pi \eta$ (channel $a$) and $K \bar{K}$ (channel $b$).
Model parameters are chosen to demonstrate key resonance features rather than to reproduce experimental results.
In this numerical example we have chosen $(m_{\rm res}, \gamma_0, g_a) = (0.984, 8, 0.75)$ (in appropriate units of GeV). 
This corresponds to a resonance width of $120$ MeV at $\sqrt{s}=m_{\rm res}=984$ MeV.

Furthermore, we examine the effective and standard spectral functions of the model, defined as~\cite{Lo:2017sde}

\begin{align}
  \begin{split}
    \label{eq:bna}
    B(s) &= 2 \frac{d}{d \sqrt{s}} \mathcal{Q}(s) \\
    A(s) &= \frac{-2 \sqrt{s} \, \sin 2 \mathcal{Q}}{s-m_{\rm res}^2}.
  \end{split}
\end{align}

\noindent The effective spectral functions within a channel can be computed via

\begin{align}
  \begin{split}
    B_i(s) &= \left[ \frac{d}{d \sqrt{s}} \,  {\rm Im}   \, \ln \, S \, \right]_{ii} \\
    &= \frac{1}{2} {\rm Im}  \left[ \, S^{-1} \frac{d}{d \sqrt{s}}  S - \left(\frac{d}{d \sqrt{s}} S^{-1} \right) S \, \right]_{ii},
  \end{split}
\end{align}

\noindent where $[ \ldots ]_{ii}$ denotes the $i-$th diagonal matrix element of a matrix. 
The standard spectral functions can be obtained directly or via the branching fractions in Eq.~\eqref{eq:branching}

\begin{align}
  \label{eq:example3}
  \begin{split}
    A_i(s) &= -2 \sqrt{s} \, {\rm Re} \, ( T_{ii} ) / (s-m_{\rm res}^2) \\
           &= \eta \times \frac{-2 \sqrt{s} \, \sin 2 \mathcal{Q}_i}{s-m_{\rm res}^2} \\
           &= {\rm BR}_i \times A(s).
  \end{split}
\end{align}

\noindent The equivalence between all the expressions in Eq.~\eqref{eq:example3} are numerically checked.\footnote{ Note that $B = 2 \frac{d}{d \sqrt{s}} (\mathcal{Q}_a + \mathcal{Q}_b) = B_a + B_b$ but $B_{a,b} \neq 2 \frac{d}{d \sqrt{s}} \mathcal{Q}_{a,b}$. }

We briefly summarize the key features of the effective spectral function $B(s)$ and 
its differences from the standard spectral function $A(s)$, see Fig.~\ref{fig:fig2}:

(i) One observes irregularities in the $B(s)$ function at $\sqrt{s}=m_\pi+m_\eta$ and $\sqrt{s}= 2\, m_K$. 
These are integrable divergences associated with the appearance of the S-wave two-body decay channel.
It can be shown that~\cite{Friman:2015zua} these threshold effects give a finite contribution to the physical observables, 
with a strength that depends on the scattering length of the channel;

(ii) The apparent ``shift'' of the strength towards lower invariant masses of $B(s)$ compared with $A(s)$ 
is a familiar feature. This effect originates from the nonresonant scattering effect, which is not properly accounted for by $A(s)$.
The enhancement near threshold can give a substantial contribution to the soft momentum of the decay particles.
This feature has been discussed in detail in Ref.~\cite{Weinhold:1997ig};

(iii) The parametrization for the resonance decay presented in Eqs.~\eqref{eq:example1} and \eqref{eq:example2} is quite robust.
It goes beyond the usual assumption of a narrow resonance, where one replaces the $\sqrt{s} \rightarrow m_{\rm res}$ in the phase spaces $\phi_i$, 
and sometimes also in prefactor multiplying $\gamma_{\rm res}$ in Eq.~\eqref{eq:example2}.
To describe higher partial waves, one needs to incorporate 
the right angular-momentum barrier to the phase space $\phi_i$.
It is a basic approximation to the integral involved in computing the imaginary part of the self energy of the resonance~\cite{Lo:2017sde,Chung:1995dx,Friman:2015zua}

\begin{align}
  g_a^2 \phi_a \leftrightarrow \int d \phi_a \, \vert \Gamma_{{\rm res} \rightarrow a} \vert^2.
\end{align}

\noindent Note that the definition for energy dependent branching fractions in Eq.~\eqref{eq:branching} remains unchanged.

In an actual PWA, the S-matrix would include multiple resonances and the effects from the nonresonant background. 
These models are constructed to describe a wide range of experimental scattering data and their parameters fitted to data.
The S-matrix obtained is usually employed to assess the existence of a resonance, and to extract its parameters.
Here we have proposed an additional use of the S-matrix -- the determination of an effective level density.

The simple example just presented motivates a robust way to extract, 
for a coupled-channel system, an effective phase shift function 
$\mathcal{Q}(s) = \frac{1}{2} \, \rm{Im}  \left( \, \ln \, {\rm det} \,S \, \right) $,
which is the suitable generalization of a single-channel phase shift.
The phase shift function $\mathcal{Q}$ thus defined is invariant under unitary rotations of the basis states, 
while the S-matrix element, which describes individual scattering process, 
depends on the choice of basis used in the coupled-channel study.

According to the S-matrix formulation of the statistical mechanics, 
the effective spectral function, $B(s) = 2 \frac{d}{d \sqrt{s}} \mathcal{Q}(s)$, 
plays the role of an effective level density due to the interaction, 
which enters the thermodynamical description of an interacting system in the form of a virial expansion.
In the next section we apply this formulation to study the thermal system of $\vert S \vert=1$ strange baryons.

\section{The pressure of strange baryons}

In the S-matrix approach to statistical mechanics, the thermodynamic pressure can be written as a sum of two pieces

\begin{align}
	P &= P_0 + \Delta P_{\rm int.}.
\end{align}

\noindent $P_0$ is the pressure of an uncorrelated gas of particles that do not decay under the strong interaction (i.e., ground-state particles), such as pions, kaons, and nucleons:

\begin{align}
\label{eqn:pgs}
    P_0 &=  \sum_{a \in {\rm gs}} d_a \int \frac{d^3 k}{(2 \pi)^3} \times T \, \left[ \pm \ln \, (1 \pm e^{-\beta (\sqrt{p^2+m_a^2}-\mu_a)}) \right],
\end{align}

\noindent where $\mu_a = B_a \mu_B + Q_a \mu_Q + S_a \mu_S$ with $(B_a, Q_a, S_a)$ being
the baryon number, electric charge and strangeness of the particle species $a$
and $(\mu_B, \mu_Q, \mu_S)$ are the relevant chemical potentials.
The choice of Fermi-Dirac or Bose-Einstein statistics depends on the quantum numbers of the species.

The interaction contribution $\Delta P_{\rm int.}$ due to two-body scatterings involves an integral over the invariant mass $\sqrt{s}$

\begin{widetext}
\begin{equation}
\label{eqn:pressure}
   \Delta P_{\rm int.} = \frac{T}{V} (\ln Z)_{int.} \\
   \approx \sum_{ab} \sum_{f} T \int_{ m_{th}^{ab} }^\infty d \sqrt{s} \int \frac{d^3 p}{(2 \pi)^3} \, 
   \frac{1}{4 i \pi} \, {\rm tr} \left[ S^{-1} \partial S-\left(\partial S^{-1}\right) S \right ]
     \left[ \pm \ln  \left( 1 \pm e^{ -\beta \left(\sqrt{p^2 + s}-\mu_a-\mu_b \right) } \right) \right].
\end{equation}
\end{widetext}
Here $S$ is the S-matrix of the scattering process $ab \rightarrow f$ with threshold $m_{th}^{ab}$ and $\partial$ stands for the derivative with respect to $\sqrt{s}$. The sum over $f$ implies summation over all allowed final states, while the
sum over $ab$ should encompass all possible pairs of ground-state hadrons: $\pi \pi$, $\pi K$, $\pi N$, $K N$, $N N$, etc.
Here and in what follows the sums over initial and final states include both particle
and antiparticle contributions.

In this paper we are interested in the pressure of strange baryons. The dominant contribution to
the strange baryon pressure comes from $|S|=1$ sector. The contribution of the  $|S|=2$ and $|S|=3$ sectors is
significantly smaller. For example, at $T=155$ MeV $|S|=2$ and $|S|=3$ baryons contribute $20\%$ and $1.4\%$
respectively to the total strange baryonic pressure. For the calculation of the $|S|=1$ strange baryonic pressure
the most important interactions are the kaon-nucleon ($KN$)
interactions, antikaon-nucleon interactions ($\bar KN$), as well as the interactions of nonstrange pseudoscalar
mesons with hyperons, i.e., the $\pi \Lambda$, $\pi \Sigma$, $\eta \Lambda$ and $\eta \Sigma$ interactions. 
The hyperon-nucleon interactions are suppressed due to the large mass of the hyperon and the nucleon.

The $\bar KN$ scattering is known to have a lot of resonances. 
As mentioned before, a coupled channel of these resonances is needed~\cite{Fernandez-Ramirez:2015tfa,Kamano:2014zba}.
Many of the resonances that are present in $\bar K N$ scattering also couple to $\pi \Lambda$, $\pi \Sigma$, $\eta \Lambda$ and $\eta \Sigma$. 
Furthermore, some of these resonances can also decay into quasi-two-body final states such as $\bar K^* N$, $\bar K \Delta$, $\pi \Sigma^*(1385)$
and $\pi \Lambda^*(1520)$, i.e., final states that contain resonances $\bar K^*,~\Delta,~\Sigma^*(1385)$ and $\Lambda^*(1520)$. 
Based on the principle of effective elementarity~\cite{Dashen:1974jw}, 
narrow resonances such as $\bar K^*$,  $\Sigma^*(1385)$ and $\Lambda^*(1520)$ (with widths of $51$ MeV, $38$ MeV and $16$ MeV, respectively), 
can be approximately treated as stable states when calculating their contributions to the thermodynamics. 
In addition, due to the long lifetime they can interact multiple times with other stable particles 
in the medium before decaying. Such interactions may be treated as effectively two-body under the same principle. 
This is in line with the framework of isobar decomposition and is compatible with the current PWA.

Taking into account that 

\begin{equation}
\frac{1}{4 i \pi} \, {\rm tr} \left( S^{-1} \partial S -\left(\partial S^{-1} \right) S \right)=\frac{1}{\pi}\frac{ d {\cal Q} }{d \sqrt{s}}=\frac{1}{2 \pi} B(s),
\end{equation}

\noindent and in particular at $\mu_Q = 0$, the pressure of $|S|=1$ baryons can be written as 

{
\begin{widetext}
\begin{align}
  P^{|S|=1}(T,\mu_B,\mu_S)=&\left[ \sum_{a=\Lambda,\Sigma,{\Sigma^*_{1535}}} d_{IJ} p_0(M_a,T) \cosh(\beta(\mu_B-\mu_S) \right] +
p_{\rm int}^{\rm res}(T) \cosh(\beta(\mu_B-\mu_S))+p_{\rm int}^{\rm KN}(T) \cosh(\beta(\mu_B+\mu_S)),\label{PS1}\\
p_{\rm int}^{\rm res}(T)=&\: T \int d \sqrt{s} \int \frac{d^3 p}{(2 \pi)^3} \frac{1}{2 \pi} B(s) e^{-\beta \sqrt{p^2+s}}=
 \frac{1}{2 \pi}  \int d \sqrt{s} p_0(\sqrt{s},T)B(s),\\
p_0(x,T)=&\: \frac{x^2 T^2}{\pi^2} K_2(\beta x).
\end{align}
\end{widetext}
}

\noindent Here $d_{IJ}=(2 J+1)(2I+1)$ is the spin and isospin degeneracy factor and $K_2(x)$ 
is the Bessel function of the second kind.
The thresholds of different scattering processes are implicitly encoded in $B(s)$. 
We used the Boltzmann approximation in the above equation since $\beta m_{th}^{ab} \gg 1$. We also performed
the calculations without the Boltzmann approximation and found that the difference is tiny. 

The first term in Eq. (\ref{PS1}) corresponds to the free gas pressure. 
In addition to the ground-state hyperons $\Lambda$ and $\Sigma$, 
we also include the narrow resonance $\Sigma^*(1385)$ in this term, since it is not reconstructed in the current PWA. 
The $\Lambda^*(1520)$ state, on the other hand, is dynamically generated, with parameters close to those in the PDG \cite{PDG}. 
It is included in the $I=0$ ($D_{03}$) component of $p_{\rm int}^{\rm res}$ (see below).
The last term in Eq. (\ref{PS1}) corresponds to $KN$ interactions and is discussed later. Here we only note that
it has a different dependence on the chemical potentials.

To evaluate the effective phase shift function $\mathcal{Q}(\sqrt{s})$ for the coupled-channel system, 
we use a coupled PWA~\cite{Fernandez-Ramirez:2015tfa} 
by the Joint Physics Analysis Center (JPAC) Collaboration,
which computes the scattering amplitude (T-matrix) for the following partial waves: 
$S_{01},~P_{01},~P_{03},~D_{03},~D_{05},~F_{05},~F_{07}$, and $G_{07}$ for isospin zero ($I=0$) and 
$S_{11},~P_{11},~P_{13},~D_{13},~D_{15},~F_{15},~F_{17}$, and $G_{17}$ for isospin one ($I=1$) cases.
Here the second subscripts in the partial-wave labels stands for the spin ($2 \times J$).
The T-matrix, and hence the S-matrix derived, describes the coupled-channel interaction of 
a system of 16 basis Fock states. It includes such major channels as $\bar K N$, $\pi \Lambda$, $\pi \Sigma$, $\eta \Lambda$, and $\eta \Sigma$.
Furthermore, for $I=0$ the quasi two-body states like $\bar K^* N$, $\bar K \Delta$, $\pi \Sigma^*(1385)$,
i.e., final states that contain resonances
$\bar K^*,~\Delta$, and $\Sigma^*(1385)$ are also included. 
As discussed above these quasi-two-body channels are important for thermodynamics.
Moreover, 
the JPAC analysis includes dummy channels labeled as $\sigma \Lambda$ and $\sigma \Sigma$
to account for the remaining inelasticities not taken into account by the channels discussed above~\cite{Fernandez-Ramirez:2015tfa}. 
Here $\sigma$ is a fictitious meson with  the mass of two pion masses. 
In principle many of these channels should be treated as genuine three-body final states.
This, at the moment, remains a challenging task. 
Nevertheless, three-body unitarity studies are currently under development in the PWA field, 
particularly related to LQCD calculations \cite{Hansen:2014eka,Briceno:2017tce,Mai:2017vot,Doring:2018xxx}.

\begin{figure}[!ht]
\includegraphics[width=3.355in]{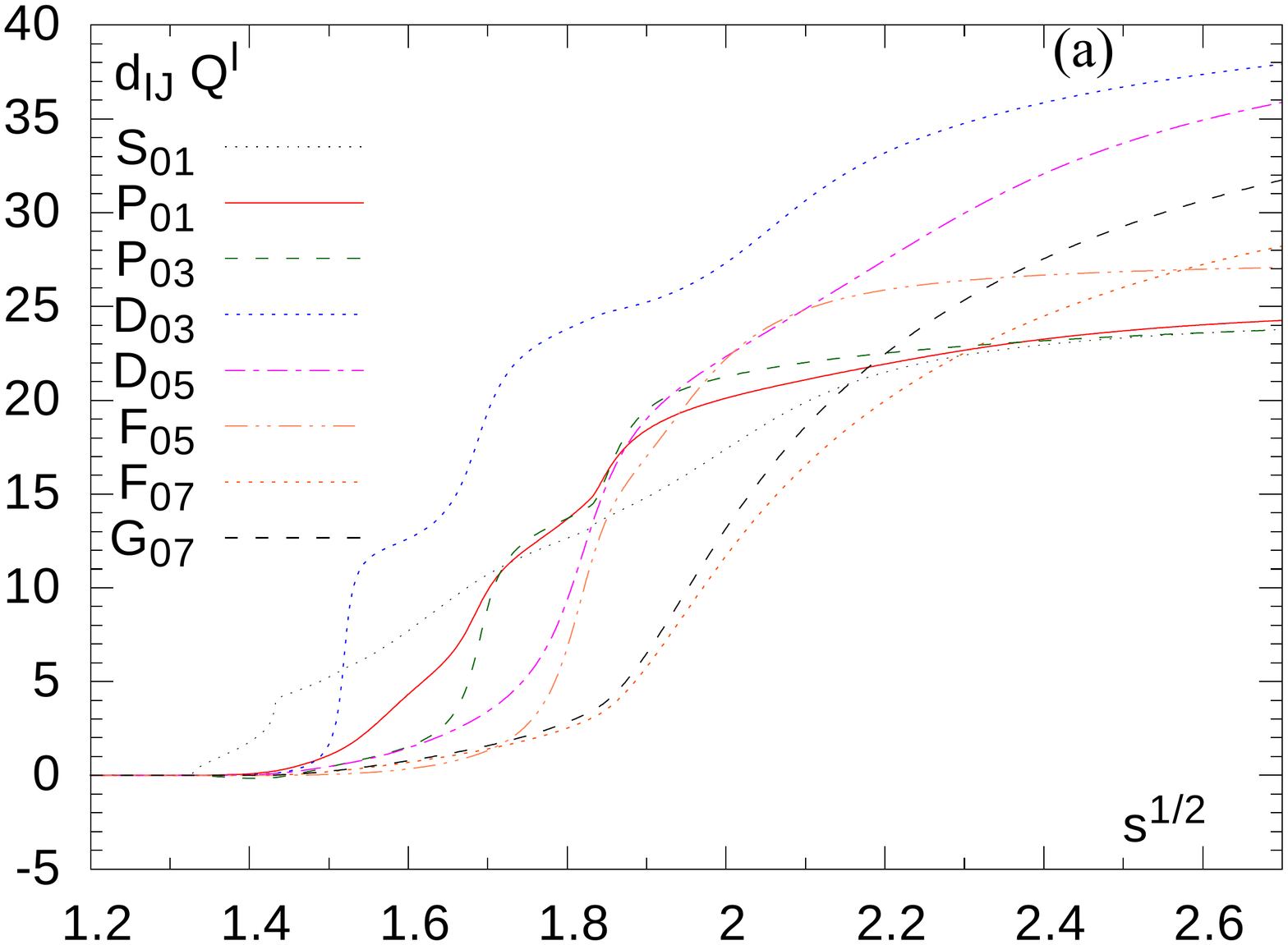}
\includegraphics[width=3.355in]{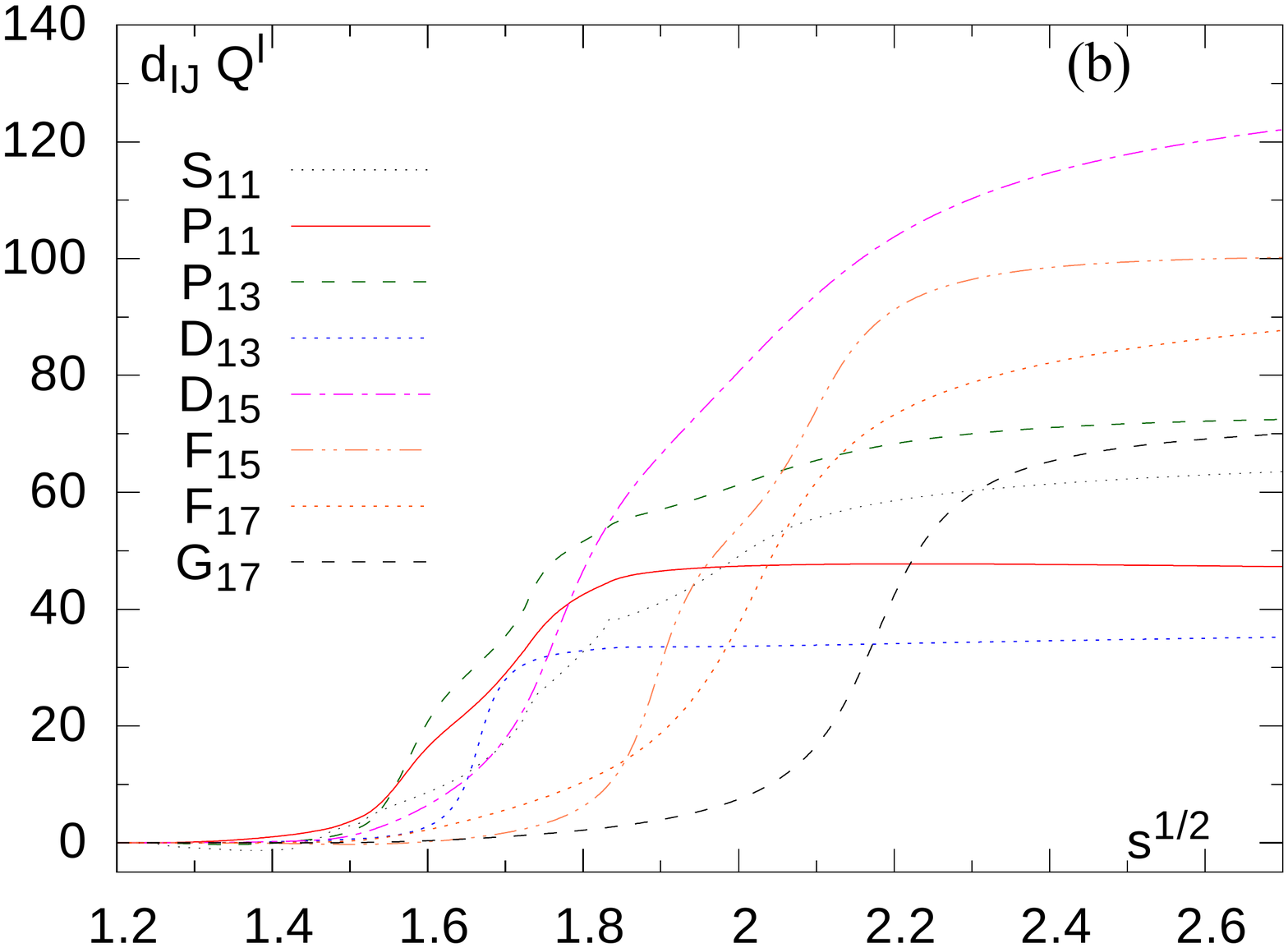}
  \caption{ 
  The generalized phase shift function $\mathcal{Q}^l(s)$ (in radians) extracted from the coupled-channel PWA in Ref.~\cite{Fernandez-Ramirez:2015tfa}. 
  Shown in the figures are the major channels contributing to the observable $\chi_{BS}$.}
\label{fig:fig3}
\end{figure}

\begin{figure*}
\includegraphics[width=8cm]{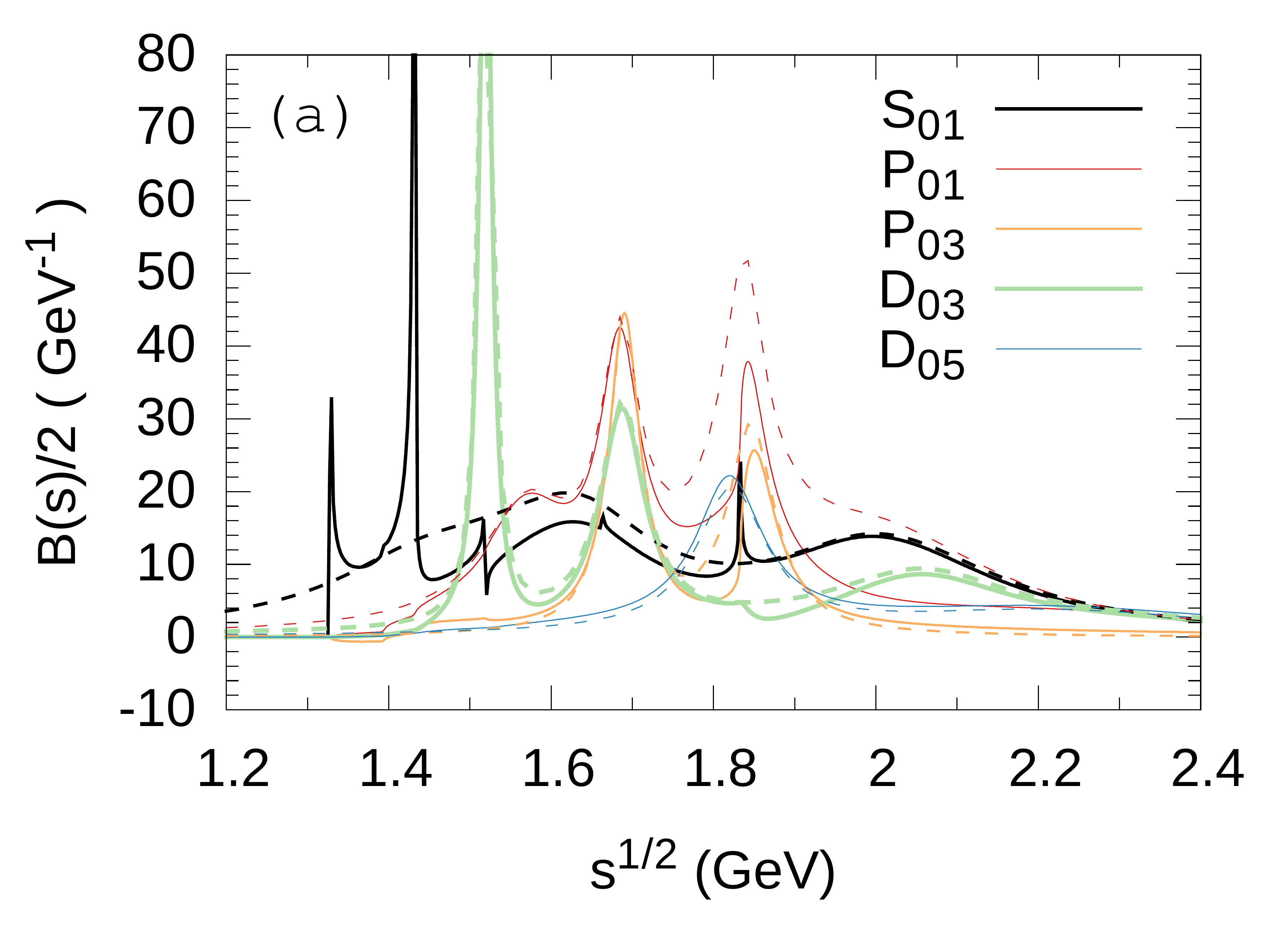}
\includegraphics[width=8cm]{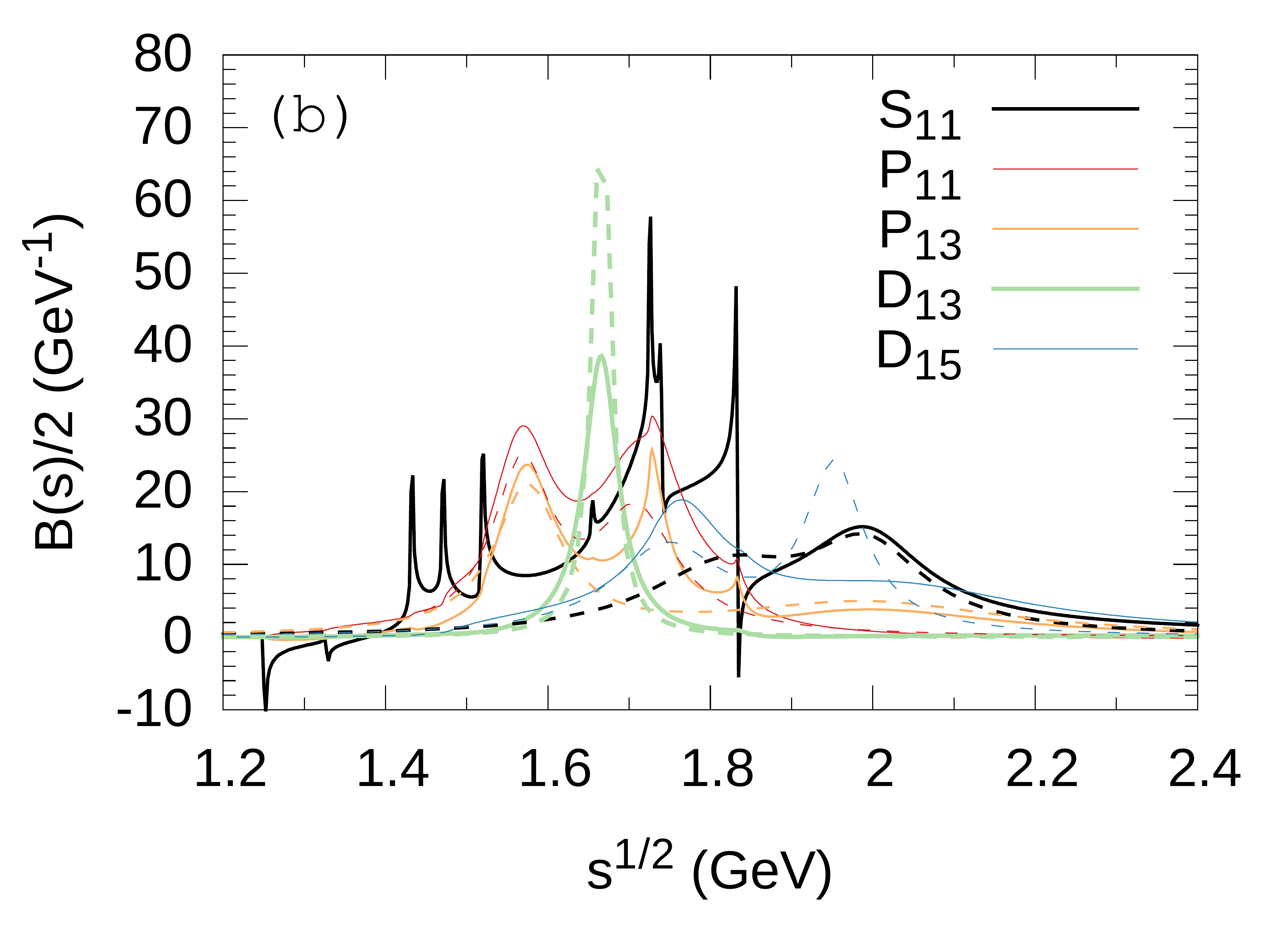}
  \caption{The effective spectral functions $B(s) \, ({\rm GeV}^{-1})$ for (a) $I=0$ and 
  (b) $I=1$ and different partial waves.
The dashed lines correspond to B-W parametrization
of the resonances.}
\label{fig:fig4}
\end{figure*}

In terms of PWA one can write
\begin{equation}
{\cal Q}=\sum_l d_{IJ} {\cal Q}^l
\end{equation}
with index $l$ labeling different partial waves, $S_{01},~S_{11},~P_{03}$, etc., and ${\cal Q}^l$ is obtained
by numerically calculating the determinant of the $16 \times 16$ S-matrix $S^l$, for each partial wave. 
The corresponding numerical results for the generalized phase shifts are shown in Fig.~\ref{fig:fig3}.
One can imagine a simple scenario where the phase shift is dominated by the sum of step functions as asserted by the HRG model.
However, Fig.~\ref{fig:fig3} shows that this simple scenario is not realized in general.
In Fig. \ref{fig:fig4} we show the effective spectral functions for the five lowest
partial waves with $I=0$ and $I=1$. We also compare the effective spectral functions
with the sum of 
B-W parametrization of the resonances in each channel. For partial waves dominated by
narrow resonances such as $P_{03},~D_{03},~P_{13}$, and $D_{13}$ the B-W parametrization gives a fair description
of the effective spectral function although not accurate at the quantitative level. The B-W parametrization also does
a fair job for the $G_{17}$ partial wave (not shown). However, for all other cases the B-W parametrization does not
describe the effective spectral function.
Furthermore, the simple K-matrix parametrization advocated in Ref.~\cite{Dash:2018can} also 
does not provide a good description of the spectral densities.
In particular, using this form
did not lead to an improved description of the spectral density compared to the simple
B-W parametrization. This is due to the fact that non-resonant background was not considered in
the K-matrix parametrization of Ref.~\cite{Dash:2018can}.
In fact, it is known that a more sophisticated treatment of the K-matrix, 
e.g., maintaining the exchange symmetry and unitarity, 
is required to produce reliable results on phase shifts and scattering amplitudes~\cite{Oller:1998hw,Doring:2006ue}.
\begin{figure}
\includegraphics[width=9cm]{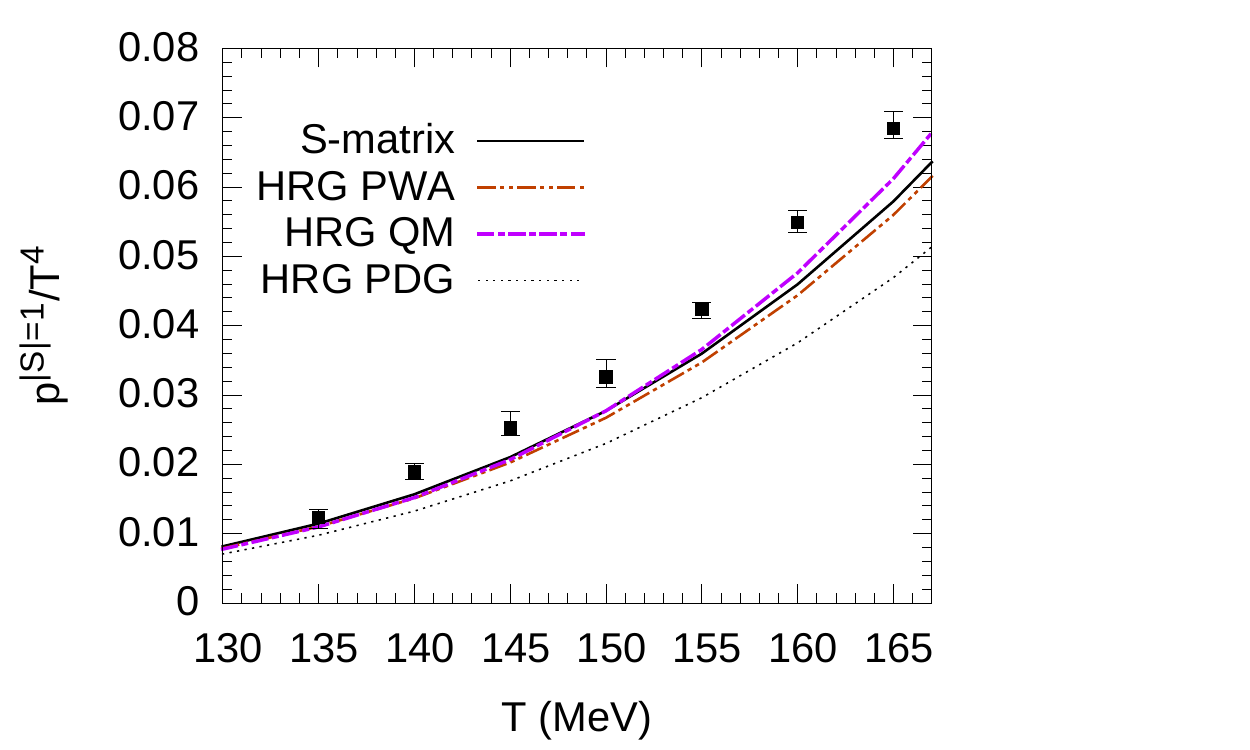}
  \caption{The pressure (normalized to $T^4$) of $|S|=1$ baryons calculated in the S-matrix based relativistic virial expansion and 
using the HRG model with various particle content 
(PWA, QM, PDG). Also shown in the figure are
the lattice results for the $|S|=1$ pressure \cite{Alba:2017mqu}.}
\label{fig:PS1}
\end{figure}

\begin{table}
\begin{ruledtabular}
\begin{tabular}{|lccc||lccc|}
  \multicolumn{4}{|c||}{$I=0$}  &  \multicolumn{4}{c|}{$I=1$}  \\
\hline
         & S-mat. & HRG   & B-W &              & S-mat.  & HRG   & B-W    \\
\hline
$S_{01}$ & 0.916  & 1.139 & 1.224  & $S_{11}$ &  1.018  & 0.282 & 0.532 \\
$P_{01}$ & 0.539  & 0.607 & 0.676  & $P_{11}$ &  1.681  & 1.275 & 1.465 \\
$P_{03}$ & 0.426  & 0.403 & 0.472  & $P_{13}$ &  1.868  & 1.857 & 2.406 \\
$D_{03}$ & 1.091  & 1.127 & 1.416  & $D_{13}$ &  0.964  & 0.995 & 1.052 \\
$D_{05}$ & 0.363  & 0.221 & 0.456  & $D_{15}$ &  1.478  & 1.219 & 1.793 \\
$F_{05}$ & 0.261  & 0.308 & 0.489  & $F_{15}$ &  0.514  & 0.503 & 1.119 \\
$F_{07}$ & 0.160  & 0.085 & 0.222  & $F_{17}$ &  0.556  & 0.238 & 0.603 \\
$G_{07}$ & 0.173  & 0.057 & 0.177  & $G_{17}$ &  0.169  & 0.095 & 0.310
\end{tabular}
\end{ruledtabular}
 \caption{The contributions to $|S|=1$ baryonic pressure from different the partial waves
in the S-matrix approach and in HRG approximation in units of $10^{-3} T^4$ at $T=150$ MeV.
In the columns labeled as ``B-W,'' the partial pressure contributions obtained
from the B-W parametrization of the spectral density are shown.}
\label{tab:tab1}
\end{table}

\begin{table*}
\begin{ruledtabular}
\begin{tabular}{|c|cccc||l|cccc|}
PW & Name & Status  & Mass (MeV)& Width (MeV) &
PW & Name & Status  & Mass (MeV)& Width (MeV)\\ 
\hline
$S_{01}$& $\Lambda(1405)$& **** & $1436$ & $279$&$S_{11}$& $\Sigma (1750)$&***&$1813$&$ 227$ \\
& ---& --- & $ 1573$ & $300$&
&$\Sigma (2000)$&*&$1991$&$173$\\
&$\Lambda(1670)$ & **** &$ 1636$ & $211$&&&&&\\
&$\Lambda (1800)$&*** & 1983 & 282&&&& &\\
& $\Lambda (2000)$&*& $ 2043$ & $350$&&&&&\\
\hline
$P_{01}$& $\Lambda(1600)$& *** & $1568$&$132$
&$P_{11}$& $\Sigma (1560)$&**& $1567$&$88$\\
& $\Lambda(1710)$& * & $1685$&$59$
&&$\Sigma (1660)$&***& $1708$&$122$\\
& --- &--- & $1835$&$180$&&&&&\\
&$\Lambda (1810)$&***& $1837$&$59$ &&&&&\\
\hline
$P_{03}$& ---& ---& $1690$&$ 46$&
$P_{13}$& ---&--- & $1574$&$99$\\
& $\Lambda (1890)$ &  **** & $1846$&$70$&
&$\Sigma(2080)$&**&$1980$&$429$ \\
\hline
$D_{03}$& $\Lambda(1520)$& **** & $1519$&$18$
&$D_{13}$& $\Sigma (1670)$&****& $1666$&$26$\\
&$\Lambda(1690)$& **** & $1687$&$66$&&&&&\\
&$\Lambda (2050)$&*&$2051$&$269$&&&&&\\
&$\Lambda (2325)$&*&$2133$&$1110$&&&&&\\
\hline
$D_{05}$& $\Lambda (1830)$&****&$1817$&$85$
&$D_{15}$&  $\Sigma (1775)$&****& $1744$&$166$ \\
&--- &--- &$2199$&$570$&
&---&---& $1952$&$88$ \\
\hline
$F_{05}$& $\Lambda (1820)$&****& $1817$&$85$&
$F_{15}$& $\Sigma (1915)$&**** & $1894$&$59$ \\
& $\Lambda (2110)$&***& $1931$&$189$
& &$\Sigma (2070)$&*&$2098$&$474$ \\
\hline
$F_{07}$& $\Lambda (2020)$&*& $2012$&$210$
&$F_{17}$&$\Sigma(2030)$&****& $2024$&$190$\\
\hline
$G_{07}$& $\Lambda (2100)$&****& $2080$&$217$
&$G_{17}$&$\Sigma(2100)$&*& $2177$&$156$
\end{tabular}
\end{ruledtabular}
\caption{Pole masses and widths of the 
$S=-1$ resonances obtained in the 
amplitude analysis of Ref.~\cite{Fernandez-Ramirez:2015tfa}.
$\Lambda(1405)$ is an effective state that accounts for the two pole structure found in the $S_{01}$ partial wave below the $\bar{K}N$ threshold~\cite{Roca:2013cca,Mai:2014xna}.
The status of the resonances is provided by the PDG~\cite{PDG} according to how well established a resonance is, with four stars being the highest qualification and one the lowest. An overall status $****$ or $***$ is awarded only to those resonances which are derived from analyses of differential cross sections and polarization observables, and are confirmed by independent analyses. All other resonances are assigned either a $**$ or $*$ status.
  } \label{tab:tab2}
\end{table*}

With the extracted effective spectral functions  $B(s)$ it is
straightforward to calculate numerically the contribution of
the coupled-channel interactions to the partial pressure of $|S|=1$ baryons.
The result is shown in Fig.~\ref{fig:PS1}. 
In 
Table~\ref{tab:tab1} we give the individual contributions to the $|S|=1$ baryonic 
pressure from different partial waves,
and in
Table~\ref{tab:tab2} we provide all the resonances
pole masses and widths
from the $S=-1$ sector PWA.
We also performed calculations by using simplified spectral functions that are the
sum of $\delta$ functions with peak positions corresponding to resonance location for
each partial wave. This corresponds to
the HRG approximation. To judge the validity of the HRG approximation in Table~\ref{tab:tab1} we compare
the contribution for each partial wave to the $|S|=1$ baryonic  
pressure with the result of S-matrix calculations. 
For partial waves, $P_{03},~D_{03},~P_{13},~D_{13}$, and $F_{15}$ HRG can reproduce the S-matrix
result for the pressure with accuracy of better than $10\%$.
However, in other cases
the contributions from the S-matrix virial expansion are either significantly smaller or significantly larger than
the HRG result. Qualitatively the situation is similar to the study of baryon number electric-charge correlations,
where it was also found that the relativistic virial expansion and the HRG model give quite different 
results \cite{Lo:2017lym}. Very interestingly, however, after adding the contribution from all partial waves, 
the two approaches give very similar results, as shown in Fig.~\ref{fig:PS1}. 
This illustrates that spectra with drastically different shape may produce similar temperature dependence in 
a given thermal observable, i.e., a given thermal quantity does not uniquely fix the spectrum. 
In particular, the excellent agreement of the HRG model with various lattice results 
should not be taken as a justification of the zero-width approximation in treating resonances. 
Such an assumption, in many cases, is not supported by empirical findings.
Instead, the current approach suggests multiple mechanisms are at work in the thermodynamic quantities: threshold effects, repulsive channels, coupled-channel effects, 
and the effect of averaging over many channels. 
A momentum-differential observable such as the momentum spectrum~\cite{Huovinen:2016xxq,Lo:2017sux} 
would allow one to differentiate between different models of the effective spectral functions.

We also investigated the question to what extent taking into account the width of the resonances 
{ as constants}
via {a} simple B-W parametrization, 
{ i.e., $\gamma_\text{res}=\gamma_0$ in Eq.~\eqref{eq:gammaedependence},}
leads to an improved description of the $|S|=1$ partial
pressure. Therefore, we calculated the pressure contributions of partial waves
using B-W parametrization of the spectral densities and show the corresponding results
in the fourth and eighth columns of Table~\ref{tab:tab1}.
{ We note that the use of the B-W amplitudes, 
with or without energy dependence in the widths, 
is not the correct way to go beyond the HRG because one
has to include both the resonant and nonresonant contributions in the spectral weight for consistently describing the thermodynamics, as done
within the S-matrix formulation of statistical mechanics~\cite{Weinhold:1997ig}}.
As one can see from Table~\ref{tab:tab1} including the width of the resonances via B-W parametrization 
often overestimates the corresponding contributions. The only partial waves, where B-W form of the spectral density
works within $10\%$ are $G_{07}$ 
and $F_{17}$, which are well described
by one isolated resonance.

\begin{figure*}
\includegraphics[width=8.6cm]{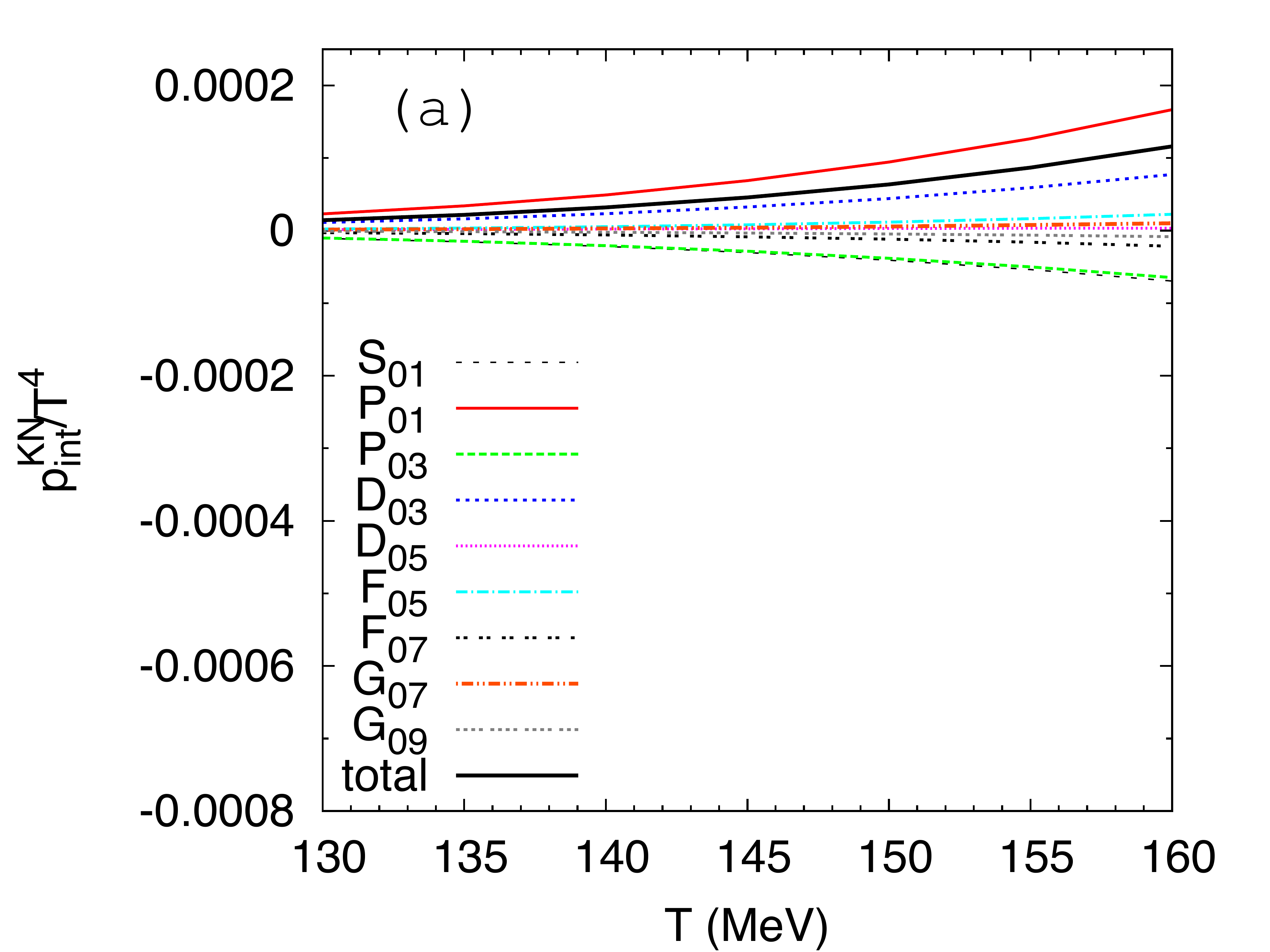}
\includegraphics[width=8.6cm]{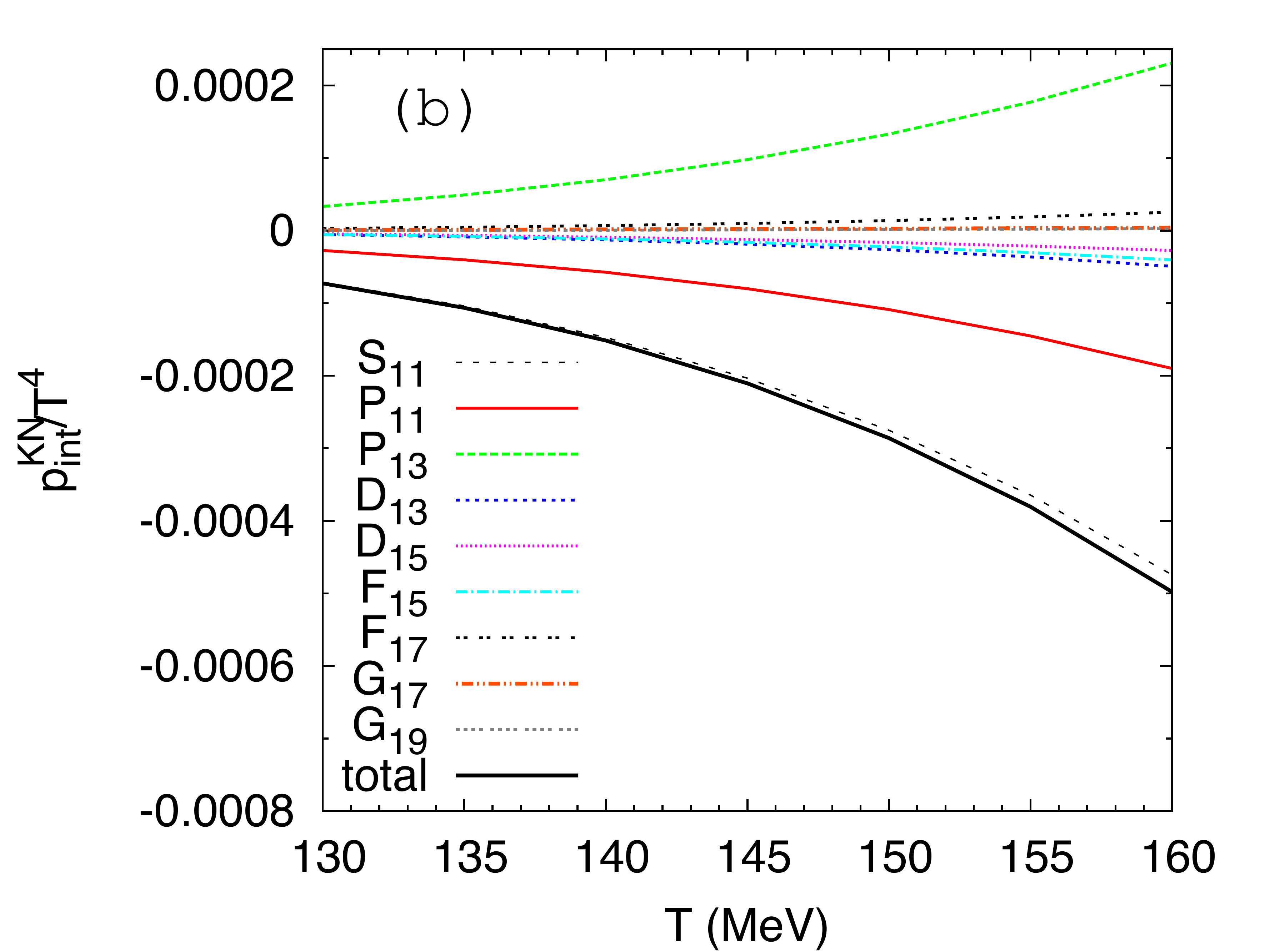}
\caption{The contribution from different partial waves to $p_{\rm int}^{\rm KN}$
  for (a) $I=0$ and (b) $I=1$.}
\label{fig:PKN}
\end{figure*}
Note that the HRG approximation discussed above and labeled as HRG PWA is different from standard
HRG model which uses only well established resonances, i.e., four and three star resonances, from PDG~\cite{PDG} and labeled as HRG PDG.
This is due to the fact that the number of hyperon resonances identified in JPAC PWA analysis 
is larger than the list of three-star or four-star resonances listed by PDG~\cite{Fernandez-Ramirez:2015tfa}.
The JPAC analysis identifies $\Lambda^*$ and
$\Sigma^*$ resonances that do not appear in PDG as well established, i.e., three-star or four-star resonances.
The analysis confirms some two- and one-star resonances that appear in PDG. At the same time other
two- and one-star resonances are not confirmed by JPAC; 
instead, new hyperon resonances are identified~\cite{Fernandez-Ramirez:2015tfa}.
Furthermore, in Fig.~\ref{fig:PS1} we show the HRG result, which in addition to the PDG states also includes 
$|S|=1$ baryons predicted in the quark model (QM) \cite{Loring:2001ky} and therefore labeled as HRG QM. 
Interestingly enough, HRG PWA and HRG QM results agree reasonably well despite differences in the particle
content. Finally, the calculations are compared with the lattice result for the $|S|=1$
baryon pressure extracted from strangeness fluctuations and baryon-strangeness correlations using
the HRG ansatz for the strange pressure \cite{Alba:2017mqu}. These lattice
results are significantly higher than the HRG result with PDG states and agree better with the 
calculations that include the missing states.

So far, we did not discuss the contribution of $KN$ interactions to the strange baryonic pressure, which
is given by the third term in Eq. (\ref{PS1}).
These interactions do not have known resonances and have been analyzed by using the
PWA in Refs.~\cite{Arndt:1984xg,Hashimoto:1984th,Hyslop:1992cs}.
The SAID interface gives the phase shift of elastic $KN$ scattering \cite{SAID}.
The inelastic channels are included explicitly through the inelasticity parameter.
To estimate the contribution of ${KN}$ we write
\begin{equation}
P_{int}^{KN}(T)=\sum_{l}  d_{IJ} \int d \sqrt{s} \,  p_0(\sqrt{s},T) 
\frac{1}{\pi} \frac{d \delta^{l}}{d \sqrt{s}}.
\end{equation}
Here it was assumed that $S=\sum_{l} \exp(2 i \delta^l)$, with
$\delta^l$ being the elastic $KN$ scattering phase shifts. We use the results of
Ref.~\cite{Hyslop:1992cs} for $\delta^{l}$ and consider
partial waves up to $G_{09}$ and $G_{19}$.
The numerical results for different partial-wave contributions to $p_{\rm int}^{\rm KN}$ are shown
in Fig.~\ref{fig:PKN} separately for the $I=0$ and $I=1$ channels. It is obvious that the largest
contribution to $p_{\rm int}^{\rm KN}$ comes from $I=1$ partial waves. 
The absolute value of   the contribution from $KN$ interactions is always smaller
than the total resonance contribution in each partial waves. The typical
strength of the $KN$ contribution relative to resonant ($\bar KN$, etc.) contribution varies between
$5\%$ and $25\%$. 
Furthermore, we observe large cancellations between the contributions from different partial 
waves; cf. Fig.~\ref{fig:PKN}. Thus, the total contribution from $KN$ interactions turns out to be very small and can be neglected.

\section{Baryon-strangeness correlations}

\begin{figure}[ht!]
 \includegraphics[width=9cm]{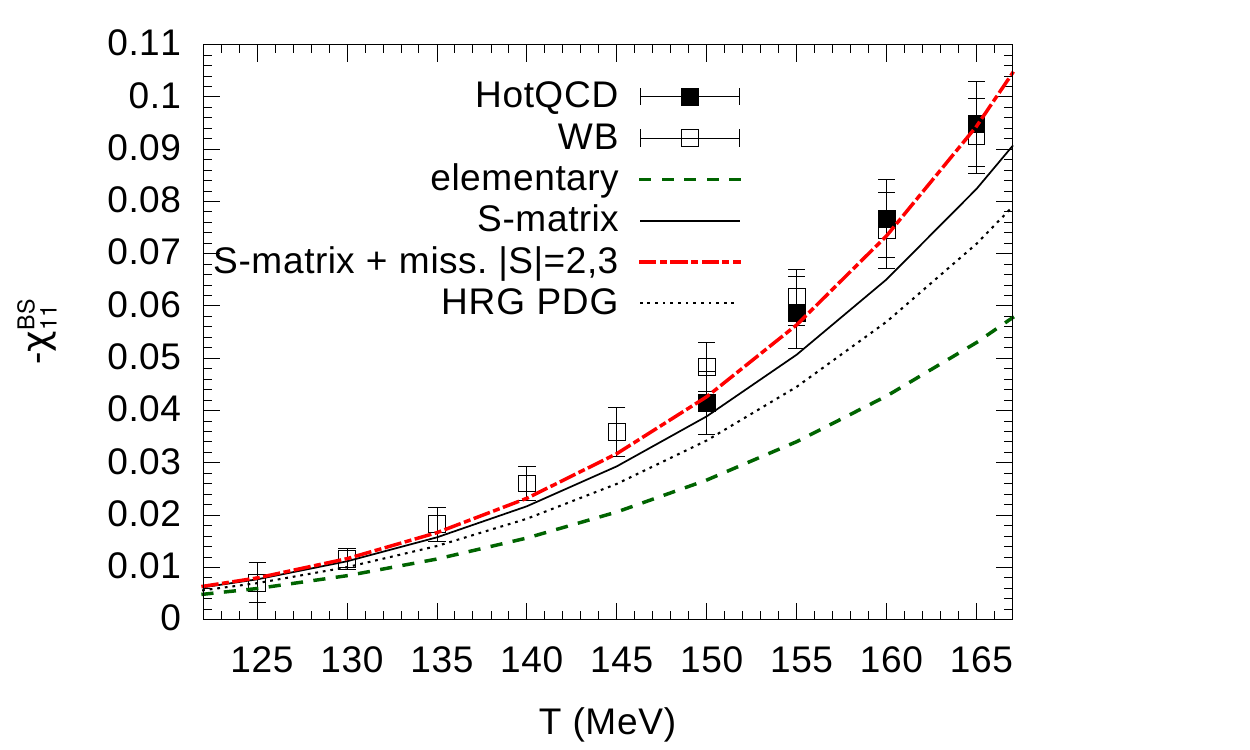}
  \caption{
The second-order baryon strangeness correlations, $\chi_{11}^{BS}$
in LQCD \cite{Borsanyi:2011sw,Bazavov:2012jq} compared with
the ideal-gas result with all elementary states (dashed line), 
with the result that includes all elementary states and interactions in $|S|=1$ sector obtained in the S-matrix
approach (solid line), as well as the result that in addition
contains the contribution from missing $|S|=2$ and $|S|=3$ states (dashed-dotted line).
Also shown is the standard HRG result (dotted line).
  }
\label{fig:BS}
\end{figure}

In the previous section we compared the pressure of $|S|=1$ baryons obtained in the S-matrix approach
with the LQCD estimate \cite{Alba:2017mqu}. The LQCD estimate of the $|S|=1$ baryonic pressure was obtained
from the baryon strangeness correlations and strangeness fluctuations and the HRG ansatz for the pressure,
and thus is model dependent.  For a more direct comparison with LQCD we consider the second-order
baryon strangeness correlation, $\chi_{11}^{BS}$. This quantity, however, receives significant contribution
from $|S|=2$ and $|S|=3$ baryons. Unlike for the $|S|=1$ sector, no PWA is available here. Furthermore, 
the number of well-established (four- or three-star) baryons is much smaller. There are only five well-established
$|S|=2$ baryons and only one well-established $|S|=3$ baryon ($\Omega(1672)$) \cite{PDG}. On the other hand, all
these states are either stable under strong interactions or quite narrow, $\Gamma \le 30$ MeV. Therefore, all
the well-established $|S|=2$ and $|S|=3$ baryons can be treated as ``elementary'' states to a good approximation, i.e.,
their contribution can be calculated by using the ideal-gas expression \cite{Dashen:1974jw}. In Fig.~\ref{fig:BS}
we show $\chi_{11}^{BS}$ obtained in LQCD \cite{Borsanyi:2011sw,Bazavov:2012jq} together with the ideal-gas result
that contains all the elementary states ($\Lambda,~\Sigma,~\Sigma^*(1385)$ and the well-established $|S|=2,3$ baryons).
The elementary states account for about $60\%$ of $\chi_{11}^{BS}$. The remainder has to come from the interactions.
If we add the interactions from the $|S|=1$ sector based on PWA, we see a substantial 
improvement in the description of the LQCD result by the S-matrix approach, over the standard HRG result that contains only the well-established states.
This demonstrates the importance of the consistent treatment of resonances and the need to incorporate additional hyperon states.

However the agreement remains incomplete, and we see that further interaction strength in the strange baryon system 
is needed to reproduce the LQCD results. This may come from an improved analysis of the $\vert S \vert =1$ hyperon system, 
e.g., more realistic interaction vertices and the systematic inclusion of the multihadron scatterings. 
For this one needs more precise experimental information on the hyperons.

The enhancement may also come from an improved treatment of the $|S|=2$ and $|S|=3$ sectors. 
Based on the quark model calculations~\cite{Loring:2001ky,Capstick:1985bm,Capstick:1986bm}, and LQCD~\cite{Edwards:2012fx}, we expect that many more $|S|=2$ and $|S|=3$ resonances should exist than in the list of well-established states in PDG.
This also suggests that interactions in the $|S|=2$ and $|S|=3$ is mostly resonant. 
To investigate the possible impact of these states we included the missing $|S|=2$ and $|S|=3$ baryon resonances to $\chi_{11}^{BS}$ by using the HRG approximation. As one can see from Fig.~\ref{fig:BS} that including these states further improves the agreement with LQCD results. 
Of course, a more definitive assessment can be made when the PWA in the $|S|=2$ and $|S|=3$ sectors becomes available.

Previously~\cite{Lo:2015cca} it was shown that the HRG model, supplemented with additional hyperon states (one- and two- star resonances), can also yield a reasonable description of the lattice results, 
despite essential differences in the distribution of strength in the effective spectral function (as a function of center-of-mass energy) between the two approaches.
This underlines the fact that spectral functions with drastically different shapes may produce similar temperature dependence in a given thermal observable. Nevertheless, the S-matrix analysis presented in this work 
is expected to yield a more reliable description since it is consistent with many known facts of the hadrons, e.g., resonance widths and inelasticities.

It would be interesting to study higher-order baryon strangeness correlations, e.g., $\chi_{22}^{BS}$ and 
$\chi_{13}^{BS}$. However, as was pointed out in Ref.~\cite{Huovinen:2017ogf} these will be sensitive
to the repulsive baryon-baryon interactions, which are not very well known in the case
of strange baryons. 
The repulsive baryon-baryon interactions need to be studied and taken into account
before applying the virial expansion to higher-order baryon strangeness correlation. 
These interactions may also explain why HRG with additional QM states seems to be disfavored
by the lattice results on the ratio $\chi_4^S/\chi_2^S$ \cite{Alba:2017mqu}.

\section{Conclusions}

In this paper the partial pressure of strange baryons and baryon-strangeness correlations have been
discussed within the S-matrix approach, based on the state of the art coupled-channel analysis by JPAC.
It was found that the proper treatment of resonances, and the natural incorporation of additional hyperon states 
which are not listed in the Particle Data Group in the S-matrix approach, 
lead to an improved description of the lattice data over the standard hadron resonance gas model. 
Thus, the analysis presented supports the earlier claim that the incorporation of extra hyperon states is required 
to explain the lattice results of the BS correlations.
Our analysis only considered meson-baryon interactions.
As we already pointed out, however, baryon-baryon interactions may be important 
for the analysis of higher-order baryon-strangeness correlations \cite{Vovchenko:2016rkn,Vovchenko:2017zpj,Huovinen:2017ogf}.

\acknowledgments

C.F.R. was partly supported by  PAPIIT-DGAPA (UNAM, Mexico) Grant No.~IA101717 and
CONACYT (Mexico) Grant No.~251817.
P.M.L. was partly supported by the Polish National Science Center (NCN), under Maestro Grant No. DEC-2013/10/A/ST2/00106 
and by the Short Term Scientific Mission (STSM) program under COST Action CA15213 (Ref. no.:41977).
P.P. was supported by U.S. Department of Energy under Contract No. DE-SC0012704. 
We used the web interface of the DATA analysis 
Center at George Washington University (http://gwdac.phys.gwu.edu/) and
Joint Physics Analysis Center (http://www.indiana.edu/~jpac/index.html) for the partial-wave analysis.
P.M.L. is grateful for the fruitful discussions with Bengt Friman, Krzysztof Redlich, Ted Barnes, Michael D{\"o}ring, and Jose R. Pel\'aez.
He also thanks Olaf Kaczmarek, Christian Schmidt and Frithjof Karsch for constructive comments 
and the warm hospitality at the Bielefeld University.
P.P. thanks Igor Strakovsky for useful correspondence and Claudia Ratti for useful correspondence and the lattice data.

\bibliography{ref}
\end{document}